\documentclass[aps,reprint,
amsmath,
amssymb,
prb,
onecolumn,
superscriptaddress,
nofootinbib]{revtex4-2}
\usepackage{dcolumn}
\usepackage{bm}
\usepackage{upgreek}

\newcommand{\bea}{\begin{eqnarray}}
\newcommand{\eea}{\end{eqnarray}}
\newcommand{\beq}{\begin{equation}}
\newcommand{\eeq}{\end{equation}}

\usepackage[dvipsnames]{xcolor}
\usepackage{mystyle}

\begin{document}

\title{
Magnon-Mediated Superconductivity in a 2D Itinerant Ferromagnet\texorpdfstring{\\}{ } with Weak Easy-plane Magnetic Anisotropy
}
\author{Vladimir Calvera }\email{vcalvera@umn.edu}- 
\affiliation{School of Physics and Astronomy and William I. Fine Theoretical Physics Institute,
University of Minnesota, Minneapolis, MN 55455, USA}
\author{Heqiu Li} 
\affiliation{Donostia International Physics Center, P. Manuel de Lardizabal 4, 20018 Donostia-San Sebastian, Spain}
\author{Yijie Wang} 
\affiliation{International Center for Quantum Materials, School of Physics, Peking University, Beijing 100871, China}
\author{B. Andrei Bernevig} 
\affiliation{Department of Physics, Princeton University, Princeton, New Jersey 08544, USA}
\affiliation{Donostia International Physics Center, P. Manuel de Lardizabal 4, 20018 Donostia-San Sebastian, Spain}
\affiliation{IKERBASQUE, Basque Foundation for Science, Bilbao, Spain}
\author{Andrey V. Chubukov }
\email{achubuko@umn.edu}- \affiliation{School of Physics and Astronomy and William I. Fine Theoretical Physics Institute,
University of Minnesota, Minneapolis, MN 55455, USA}
\begin{abstract}
Motivated by recent observations of superconductivity in a quarter-metal state of spin- and valley- polarized graphene multilayers, we investigate pairing within a ferromagnetic phase of a single-valley model of itinerant two-dimensional (2D) electrons with Hubbard-type interaction and no artificial high-energy cutoff. 
In 2D, the Stoner transition is first-order into a fully-polarized state wherein the only gapless collective excitations are transverse magnons. 
We find that in a spin-SU(2) symmetric model, this magnon-mediated pairing interaction between equal-spin fermions vanishes at $T=0$.
We show that a small easy-plane magnetic anisotropy $\Omega_0 \ll E_F$, where $E_F$ is 
the Fermi energy, breaks the SU(2) symmetry and generates an attractive interaction for equal-spin $p-$wave pairing.
We explicitly derive the corresponding coupling constant $\lambda_p$ as the scaling function of both the relative strength of the easy-plane anisotropy, $\Omega_0/E_F$, and the proximity to the ferromagnetic transition. While $\lambda_p$ is parametrically small in $\Omega_0/E_F$ deep inside the ferromagnetic phase, it becomes enhanced near the ferromagnetic transition, reaching order unity regardless of how small $\Omega_0/E_F$ is. This mechanism yields a sizable $T_c$, peaked near the onset of ferromagnetism.
\end{abstract}

\maketitle

\section{Introduction}
\label{sec:Intro}
Non-phononic mechanisms  of superconductivity continue to attract a high level of attention and interest in the condensed matter community. A positive semi-definite (repulsive) electron-electron interaction has partial components in all pairing channels; for a rotationally-invariant system, when the Fermi surface screening is accounted for, the interaction stops being positive semi-definite and acquires attractive channels, as Kohn and Luttinger first demonstrated~\cite{kohn_luttinger}. Pairing  by magnetic fluctuations -- an extension of Kohn-Luttinger mechanism-- remains a viable candidate for almost all novel superconductors, from $^3\rm{He}$ \cite{layzer1971spin, Leggett1975ReviewHe3, fay1980coexistence,Vollhardt_2013} to cuprates and Fe-based~\cite{Scalapino_2012}, and has been discussed in the context of superconductivity in twisted and non-twisted graphene-based materials and transition metal dichalcogenides~\cite{Zhou2021,Zhou2021a,balents2020superconductivity,Zhang2023a,Patterson2024,Holleis2025,kim2026resolving}.
 
The subject of this communication is pairing mediated by ferromagnetic fluctuations near the onset of a ferromagnetic order. 
Such a pairing has been studied by many groups in the paramagnetic phase near the transition~\cite{layzer1971spin, Monthoux1999,Wang2001,Roussev2001,Chubukov2003,Rosch2007,Chubukov2020a}.
Recent discoveries of superconductivity in a half-metal state of Bernal bilayer graphene (BBG) and rhombohedral tri-layer graphene (RTG) and in a quarter-metal state in pentalayer graphene call for understanding of superconductivity located within a ferromagnetic state: a quarter metal has both valley and ferromagnetic order~\cite{Zhou2021} and magnetometry measurements in BBG and RTG identified a half-metallic state there as a ferromagnet. 
Superconductivity inside a ferromagnetic state has been earlier detected in 3D heavy fermion systems such as UGe$_2$~\cite{Saxena2000}, URhGe~\cite{Aoki2001} and UCoGe~\cite{Huy2007}, and is also present in magic angle twisted bilayer graphene as the system exhibits what are thought to be strongly correlated magnetic insulators~\cite{balents2020superconductivity,Andrei2021Marvels}. 
   
From a theoretical perspective, magnetically-mediated pairing in the ferromagnetic phase in 3D has been studied in Refs.~\cite{fay1980coexistence,Kirkpatrick2001}. There are two candidates for a pairing boson: massless transverse (Goldstone) magnon excitations and massive longitudinal excitations. The authors of Ref.~\cite{fay1980coexistence} assumed that the pairing is mediated by longitudinal spin fluctuations, and the authors of Ref.~\cite{Kirkpatrick2001} further argued that such pairing is enhanced due to coupling between longitudinal and transverse excitations.
   
Pairing mediated by longitudinal spin fluctuations is an established scenario for 3D itinerant ferromagnets\cite{Fay1980,Kirkpatrick2001,Kirkpatrick2003}. In these systems, the Fermi surfaces for both spin components coexist below the transition temperature, albeit with unequal sizes, allowing strong longitudinal magnetic fluctuations to act as a pairing glue.
In contrast, the situation is fundamentally different in graphene-based 2D systems. Quantum oscillation data reveal that a ferromagnetic  transition is first order into a half-metal with a Fermi surface for only a single spin projection. This leaves a transverse magnon with particle-like dispersion $\omega \propto q^2$ in the SU(2)-symmetric case as the only electronic candidate for a pairing glue.~\footnote{Some data are consistent with a Partially Isospin polarized (PIP) state, which is "almost" a  half-metal, but there are still small Fermi surfaces for minority carriers. In this situation, longitudinal fluctuations are present, but are severely reduced.}
 
Pairing by magnons in 2D ferromagnets has been studied recently for both two-valley and one-valley systems~\cite{Dong2024,Dong2025,Raines2026,raines2025superconductivity}. In this paper, we focus on the one-valley case with the ultimate goal to understand superconductivity in a quarter-metal. We consider 2D fermions with parabolic dispersion $k^2/(2m)$ and Hubbard-type interaction $U$  within Hartree-Fock (ladder) approximation.  Within this approximation, the strength of fluctuations leading to a Stoner  ferromagnetism is measured by a dimensionless quantity 
\begin{equation}\label{eq:c:definition}
    c \equiv \frac{Um}{2\pi}.
\end{equation} 
At $c =1$, the system undergoes a 
first-order transition from a paramagnet  at $c<1$ to a fully polarized  ferromagnet at $c>1$.
This transition is rather unconventional as 
 the spin susceptibility of the paramagnet diverges at $c \to 1$ from below. As a consequence, in a half-metal state near the transition, magnons are heavy quasiparticles:  their effective mass 
 $m$ diverges as $1/(c-1)$.
 
The discussion of the pairing within a magnetically ordered state is often centered around the issue of a reduction of the pairing vertex mediated by a Goldstone boson because the coupling between a Goldstone boson and low-energy fermions must vanish at vanishing bosonic momentum and frequency  to preserve the form of the bosonic propagator
(in high-energy this is known as  Adler principle~\cite{Adler1965}).
This supression
is the primary reason why $d$-wave pairing, mediated by antiferromagnetic spin fluctuations, is strongly suppressed once the system develops long-range antiferromagnetic order~(see, e.g., \cite{Schrieffer1995}).
In this communication we argue that there is a far stronger no-go argument against magnon-mediated pairing in a saturated ferromagnet with Fermi surface for only spin-up fermions. Namely, in the SU(2)-invariant model, the effective pairing interaction between spin-up fermions strictly vanishes at $T=0$. This vanishing occurs because the pairing interaction is a convolution of two magnon propagators and two propagators of gapped spin-down fermions; since all the poles of these Green's functions lie in the same complex half-plane, the internal frequency integral vanishes after closing the contour in the other half-plane. 
The crucial element here is the fact that in an SU(2) symmetric model a magnon propagator has a single pole. 
Another intuitive picture is to instead go to imaginary time $\tau$ and notice that in time domain the effective pairing interaction  can be regarded  as a circular process in which a fermion has to return to the same $\tau$ at which it started, yet  at each step of the circular process it undergoes a  time delay.

The authors of Ref.~\cite{raines2025superconductivity} by-passed this no-go argument by adding  phenomenologically an upper frequency cutoff $\Lambda$ to the magnon propagator. They speculated that this cutoff comes from other bands, which do not cross the Fermi level and for this reason do not affect the development of ferromagnetism and magnon-mediated pairing, but affect the physics at high energies. They argued that once the frequency integral does not vanish,  there is an attractive interaction for  equal spin pairing. The attraction is reduced, consistent with the Adler principle, but remains non-zero. The  drawback of this consideration is that to get the sizable magnitude of the attractive {$p$-}wave coupling  
\begin{equation}\label{eq:lambdap:Intro}
    \lambda_{p} \equiv -\int_0^{\pi} \Gamma(\phi)\cos(\phi) \frac{\dd\phi}{\pi},
\end{equation}
where $\Gamma (\phi) $ is the dimensionless pairing interaction between fermions with momenta $({\bf k}_F, -{\bf k}_F; {\bf p}_F, -{\bf p}_F)$ and ${\phi}$ is the angle between 
$ {\bf k}_F$ and $ {\bf p}_F$, the authors of \cite{raines2025superconductivity} had to assume that the upper cutoff $\Lambda$ is comparable to the Fermi energy $E_F$, otherwise (for larger $\Lambda$), $\lambda_p$ becomes parametrically small in $E_F/\Lambda$.   
  
The issue we address here is whether a pure single-band model of 2D ferromagnetically ordered fermions with no phenomenologically imposed cutoff can still undergo magnetically-mediated {$p$-}wave superconductivity at a reasonably  high temperature.  We argue that this does occur if there is a small easy-plane magnetic anisotropy induced by spin-orbit coupling.  We assume that this anisotropy modifies the magnon propagator at energies below $\Omega_0 \ll E_F$. At a first glance, $\lambda_p$ should be parametrically small in $\Omega_0 (\ll E_F)$. In Sec.~\ref{sec:SCCouplingConstant}, we argue that this is indeed the case deep inside a ferromagnetic phase, where $\lambda_p \propto (\Omega_0/E_F)^{3/2}$ is obviously small. 

However, the situation is different close to the onset of ferromagnetism, when $c-1$ is also small. We show explicitly that in this range 
\begin{equation}
    \lambda_{p} = \frac{c}{c-1} F\left((c-1) \frac{\EF}{4c\Omega_0 }\right),
\label{ch_1}
\end{equation}
where $F(z)$ is a {\it universal}  function of the argument $z = (c-1)E_F/(4c \Omega_0)$ (See Eq.~\ref{eq:ScalingFunctionFofz}). 
At $z\gg 1$, where our  calculations are under control, we find that $F(z)$ is positive and scales as $F(z) \approx 0.1/z^{3/2}$.  
The positive $\lambda_p$ then behaves as $\lambda_p \propto 0.1/((c-1) z^{3/2})$. It rapidly increases as $c-1$ decreases  because of $1/(c-1)$ in the prefactor and because $z$ gets smaller, and becomes unity at $(c-1)\approx  0.4 (\Omega_0/E_F)^{3/5}$.  At even smaller values of $c-1$, $\lambda_p$ becomes larger than one.  
In this  situation, we expect that the frequency dependence of the pairing interaction becomes relevant and likely keeps  $\lambda_p = \Omc(1)$  (see e.g.,  Ref.~\cite{Abanov2020}). 
At even smaller $c-1 < 4 \Omega_0/E_F$,  $z$ becomes smaller than one. Our result for $F(z)$ for these $z$ is not fully controllable (corrections are $\Omc(1)$),  but taken at a face value shows that $F(z)$ start decreasing and  eventually changes sign from attractive to repulsive. 
The outcome is that superconducting $T_c$ for equal-spin {$p-$}wave pairing is peaked within the ferromagnetic phase, at a small distance from  its onset, at $c-1 \sim (\Omega_0/E_F)^{3/5}$. It decreases when the system moves deeper into the ferromagnetic region and also when it comes closer to a transition to a paramagnet.

The paper is organized as follows. In the next Sec.~\ref{sec:Model} we introduce the one-band model with parabolic dispersion and Hubbard-type interaction and discuss Stoner transition to ferromagnetism, magnon spectrum and the effect of a small easy-plane anisotropy. In Sec.~\ref{sec:EffectiveInteractions} we consider an effective electron-magnon interaction and four-fermion interaction mediated by two magnons. In Sec.~\ref{sec:SCCouplingConstant} we obtain the coupling in the {$p$-}wave superconducting channel. We present our conclusions in Sec.~\ref{sec:conclusions}

\section{Model}\label{sec:Model}

\subsection{Isotropic model}
We consider the model of an isotropic itinerant ferromagnet  with $k^2/(2m)$ dispersion and Hubbard-type 4-fermion interaction. 
 The model is described by Matsubara action
\begin{equation}
\begin{split}
    S = &\sum_{\underline{k};\sigma} \bar{\psi}^{\,}_{\underline{k}\sigma}
    \left(
    -\ii\omega +\frac{\kbs^2}{2m}-
    \mu_0
    \right)\bar{\psi}^{\,}_{\underline{k}\sigma}
    + 
    \frac{U}{2}\sum_{\underline{k}\underline{k}'\underline{q};\sigma\sigma'}
    \bar{\psi}_{
    \underline{k}+\underline{q}\sigma
    }
    {\psi}_{
    \underline{k}\sigma
    }
    \bar{\psi}_{
    \underline{k}'-\underline{q}\sigma'
    }
    {\psi}_{
    \underline{k}'\sigma'
    }
    \end{split},
\label{aa:1}
\end{equation}
where $\underline{k} = (\kbs,\ii\omega)$ combines fermionic momentum and frequencies, and $\sigma \in \{\ua,\da\}$ denotes spin. 
The interaction strength is measured in terms of the dimensionless parameter
\begin{equation}
    c \equiv \nu U,
\end{equation}
where $\nu = \frac{m}{2\pi}$ is the density of states per spin. We work at fixed electron density $\nel$, which defines a momentum scale $\kF\equiv \sqrt{4\pi  n_e}$ and energy scale $2\mu_0\equiv\frac{\kF^2}{2m}$. As we are working in the fully ferromagnetic phase, the Fermi energy is always given by $\EF = \frac{\kF^2}{2m}$.

To detect the Stoner transition, we restrict to the ladder approximation\cite{shimizu1981itinerant,Raines2024a,Raines2024b,Raines2026}. 
We introduce a trial infinitesimally small ferromagnetic order parameter $\Delta_0$ and compute the fully dressed order parameter, which we label as $\Delta$, by summing up ladder diagrams in the particle-hole channel. The ratio $\Delta/\Delta_0$, which is proportional to the ferromagnetic susceptibility, diverges at $c=1$, signaling the onset of a ferromagnetic order. 
For $c >1$, we introduce a spontaneous non-zero $\Delta$ without $\Delta_0$, adjust the Green's functions for spin-up and spin-down fermions and solve the non-linear self-consistent equation on $\Delta$. We find that the solution $\Delta (c)$ jumps at $c =1+0^+$ to a finite value 
 \begin{equation}
    \Delta \equiv U \expval{n_{\ua}-n_{\da}} =  c\mu_0.
\end{equation}
One  can obtain the same result by calculating the energies of the paramagnetic fluid and ferromagnetic fluid within the Hartree-Fock approximation. In the ferromagnetic state, the spin-up and spin-down electron Green's functions for this $\Delta$ are given by
\begin{equation}
   \begin{split}
        G_{\ua}(\ii\Omega,\kbs) &= \frac{1}{\ii\Omega- \frac{\kbs^2}{2m} + 2\mu_0},\\
    G_{\da}(\ii\Omega,\kbs) &= \frac{1}{\ii\Omega- \frac{\kbs^2}{2m} + 2(1-c)\mu_0},
   \end{split}
\end{equation}
We see that spin-up fermions have a Fermi surface, but spin-down fermions are gapped when $c>1$.
Because all the spin-down electrons are gapped $(n_\da=0)$, the system is a half metal; consequently, $\Delta$ attains its largest possible value.

A magnon is a Goldstone mode associated with magnetic fluctuations transverse to the magnetization, whose expectation value we choose to be along $z-$direction. 
Consequently, the magnon appears in correlation functions of the transverse spin densities $\hat{S}^{x/y}=\sum_{\kbs,\omega}\bar{\psi}_{\ii\omega,\kbs}\sigma^{x/y} \psi_{\ii\omega,\kbs}$.
As the FM order parameter is invariant under rotations generated by $\hat{S}_z$, it is convenient to look at correlation functions of the operators 
$\hat{S}^\pm := \hat{S}^x \pm \ii \hat{S}^y$, which can be equivalently expressed as ${\hat S}_{\ua\da}$ and ${\hat S}_{\da\ua}$. The corresponding spin-flip susceptibility is, within the ladder approximation,   
\begin{equation}\label{eq:chi:ud}
    \chi_{\ua\da}(\ii\Omega,\qbs) = \frac{\Pi_{\ua
\da}(\ii\Omega,\qbs)}{1-U \Pi_{\ua
\da}(\ii\Omega,\qbs)}.
\end{equation}
where the bare spin-flip polarization bubble $\Pi_{\ua\da}$ is 
\begin{equation}\label{eq:SpinFlipSusceptibilitity}
    \Pi_{\ua\da}(\ii\Omega,\qbs) \equiv -\int_{\ii\omega,\lbs}
    G_{\ua}(\ii[\Omega+\omega],\qbs+\lbs)
    G_{\da}(\ii[\omega],\lbs).
\end{equation}
 The analytical expression for $  \Pi_{\ua\da}(\ii\Omega,\qbs) $ is 
\begin{equation}
    \Pi_{\ua\da}(\ii\Omega,\qbs) =\frac{m}{2\pi} \frac{m}{q^2}\left(\frac{q^2}{2m}+2\Delta + \ii\Omega  - \sqrt{\left(\frac{q^2}{2m}+2\Delta + \ii\Omega \right)^2 - \frac{4\Delta}{c}\frac{q^2}{m}}\right).
\label{aa:2}
\end{equation}
where $q := \abs{\qbs}$  (see  App.~\ref{app:SpinSusceptibilityANDMagnonPole} for details).
Substituting into Eq.~\ref{eq:chi:ud} and performing analytical continuation to the upper half-plane ($\ii\Omega \to \omega + i\delta$), we find that $\chi_{\ua\da}(z,\qbs)$ has a magnon pole at $\omega = -\alpha \qbs^2 - \ii 0^+ $, where the `spin stiffness' is 
\begin{equation}\label{eq:SpinStiffness}
    \alpha = \frac{c-1}{2mc}.
\end{equation}
Near the pole, 
\begin{equation}
  \chi_{\ua\da}(\ii\omega,\qbs)  \approx \frac{2\Delta}{U} \frac{Z(\qbs)}{\omega + \ii 0^+ + \alpha q^2}.
\end{equation}
 The residue of the magnon pole is  
\begin{equation}\label{aa:5}
   Z (\qbs)= \left(1 - \frac{\qbs^2}{q^2_c}\right); \quad  (\abs{\qbs}^2 \leq q^2_c).
\end{equation} 
where  $q_c = \sqrt{4mc\Delta} = c k_F$. 
At $\abs{\qbs}=0$,   $Z(\zero) =1$.  As $\abs{\qbs} $ increases,  $Z (\qbs)$  gets smaller  and eventually vanishes at  $\abs{\qbs} = q_c$.   At larger $\abs{\qbs}$, there is no magnon pole.
  \begin{figure}
    \centering
    \includegraphics[width=0.75\linewidth]{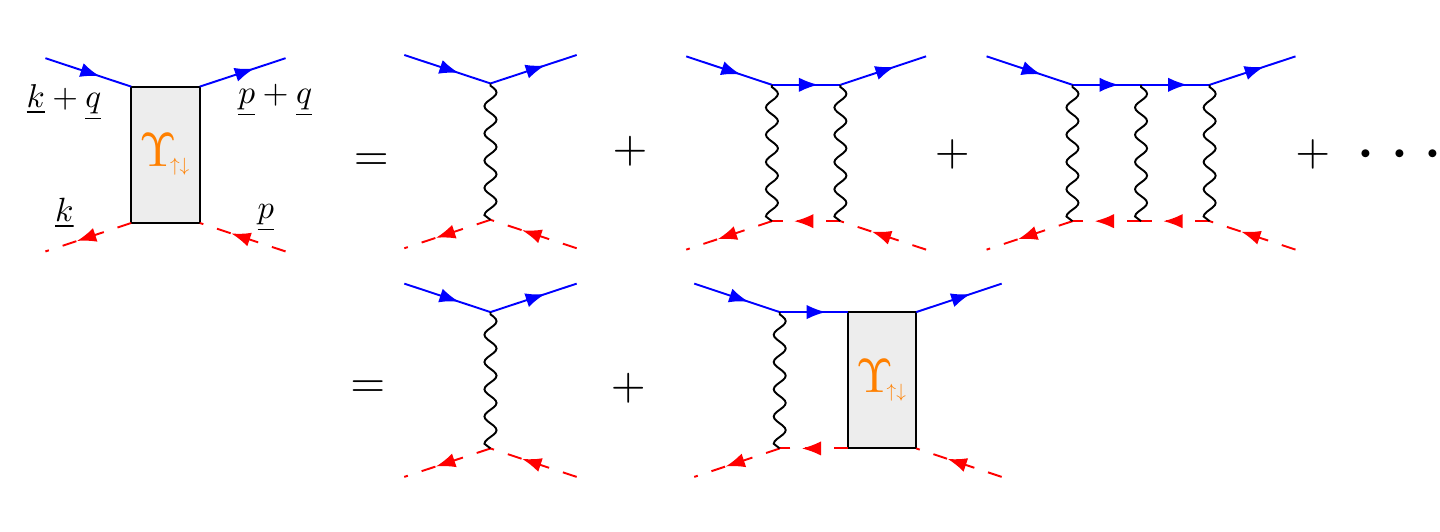}
    \caption{The effective interaction ($\Upsilon_{\ua\da}$) between spin-up (blue solid lines) and spin-down electrons (red dashed lines)
     in the Ladder approximation. The Wavy lines denote the bare interaction and the underlined symbols denote frequency-momentum vectors, e.g., $\underline{k}= (\ii\omega ,\kbs)$. 
}
    \label{fig:LadderSeries:Upsilon}
\end{figure}

\begin{figure}
    \centering
    \includegraphics[width=0.4\linewidth]{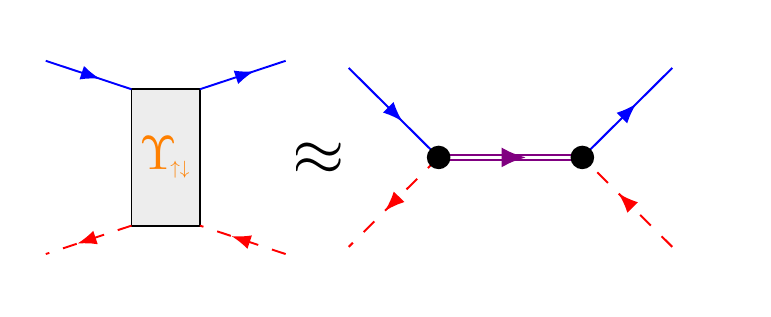}
    \caption{Magnon-propagator approximation: we approximate the 4-fermion interaction, given by the series of ladder diagrams in Fig.~\ref{fig:LadderSeries:Upsilon},  as the effective interaction mediated by a magnon.  Double (purple) line denotes the magnon propagator, and black dots represent electron-magnon vertices. Blue and red solid lines denote spin-up and spin-down electrons, respectively. }
    \label{fig:MagnonPole:Approximation}
\end{figure}

We next  obtain the effective dynamical electron-electron interaction $\Upsilon_{\ua\da}(\ii\Omega,\qbs)$, mediated by a magnon. 
This effective interaction is obtained from the Hubbard $U$ by adding an infinite series of  ladder diagrams that contain the same polarization 
$ \Pi_{\ua\da}(\ii\Omega,\qbs) $ as in the spin-flip susceptibility.  The ladder diagrams are shown in  Fig.~\ref{fig:LadderSeries:Upsilon}. 
For a generic interaction that depends on  momentum and/or frequency transfer, $\Upsilon_{\ua\da}$ depends on the three external momenta and frequencies, 
$\underline{k},\underline{p},\underline{q}$,
 and the second line in  Fig.~\ref{fig:LadderSeries:Upsilon}  is expressed as an integral  equation. However, for a 
 $U$,  $\Upsilon_{\ua\da}$  depends only on the transferred momentum and frequency $\underline{q}=(\ii\Omega,\qbs)$, and the ladder series reduce to the  algebraic equation 
\begin{equation}
    \Upsilon_{\ua\da}(\ii\Omega,\qbs) = U + U\Pi_{\ua\da}(\ii\Omega,\qbs) \Upsilon_{\ua\da}(\ii\Omega,\qbs).
\end{equation}
Solving this equation, we find
\begin{equation}\label{eq:MagnonPropagator:ISO}
    \Upsilon_{\ua\da}(\ii\Omega,\qbs)
     = \frac{U}{1- U\Pi_{\ua\da}(\ii\Omega,\qbs)} = U + U \chi_{\ua\da}(\ii\Omega,\qbs)U,
\end{equation}
where in the last term we re-expressed the solution in terms of the spin-flip susceptibility, given by Eq.~\ref{eq:chi:ud}. Isolating the magnon 
contribution, we find 
\begin{equation}
 \Upsilon_{\ua\da}(\ii\Omega,\qbs) \approx 2\Delta U  D_{\ua\da}(\ii\Omega,\qbs),
 \label{aa:4}   
\end{equation}
where $D_{\ua\da}$ is the magnon propagator on the Matsubara axis: 
\begin{equation}
       D_{\ua\da}(\ii\Omega,\qbs) = \frac{Z (\qbs)}{\ii\Omega+\alpha\qbs^2}.
\end{equation}
We represent this effective interaction graphically by drawing a double line to denote the magnon propagator, and black dots to represent the scattering vertex from a spin-up electron to a  spin-down electron by emitting a magnon (see Fig.~\ref{fig:MagnonPole:Approximation}). We note  that in the more general setting of momentum dependent 4-fermion  interaction, the electron-electron-magnon vertex acquires  dependencies on $\kbs$ and $\pbs$, while the magnon propagator only depends on $\underline{q}=(\ii\Omega,\qbs)$. 

\subsection{Anisotropic model}

We now introduce a small easy-plane anisotropy that favors magnetization in the $y-z$ plane and  breaks the spin $\SU(2)$ symmetry down to the $\U(1)$ subgroup generated by $\hat{S}_x$. Spontaneous ferromagnetic order now breaks only a single continuous symmetry generator, compared to two in the isotropic limit. This reduction directly alters the  structure of the magnon propagator.
As was demonstrated in \cite{watanabe2012unified}, the Goldstone modes (GMs) can be classified as either type-A or type-B. Type-A NGMs arise from a single broken generator, whereas type-B GMs are formed by a canonical pair of broken generators. This structural difference manifests in their propagators: the low-frequency expansion of the inverse propagator is $\Omega^2$ for type-A GMs and  $\Omega$ for type-B GMs. 
In our case, in the SU(2)-isotropic case, the magnetization breaks the rotations generated by $\hat{S}_{x}$ and $\hat{S}_y$. Because their commutators obey the commutation relation $[\hat{S}_x,\hat{S}_y] =\ii\hat{S}_z$ and $\expval{\hat{S}_z}\neq 0$, they form a conjugate pair, rendering the magnon a type-A NGM. In the U(1) case, an easy-plane anisotropy explicitly breaks the rotation symmetry generated by $\hat{S}_x$. The magnetization then spontaneously breaks only the remaining generator $\hat{S}_y$, forcing the Goldstone mode to be type-A.
 
To quantify the effect of the anisotropy, we introduce the scale $\Omega_0$, above which the GM is type-B, whereas below $\Omega_0$, it behaves as type-A and 
modifies the magnon propagator, $ \tilde{D}_{\ua\da}(\ii\Omega,\qbs)$, to 
\begin{equation}\label{eq:MagnonPropagator:ANI}
    \tilde{D}_{\ua\da}(\ii\Omega,\qbs) = \frac{Z(\qbs)}{B(\ii\Omega)+\alpha \qbs^2},
\end{equation}
where 
\begin{equation}
   B(\ii\Omega) := \frac{\Omega^2}{\sqrt{\Omega_0^2-(\Omega+\ii 0^+)^2}}=\begin{cases}
       \frac{\ii\Omega}{\sqrt{ 1-\Omega_0^2/\Omega^2}} \quad &, \abs{\Omega} \geq \Omega_0;\\
        \frac{\Omega^2}{\Omega_0\sqrt{1- \Omega^2/\Omega_0^2}} \quad &, \abs{\Omega} \leq \Omega_0.\\
   \end{cases}
\end{equation}
To simplify the analysis, we neglect the effect of the magnetic anisotropy on the electronic Green's functions. The effective magnon-mediated interaction between spin-up and spin-down fermions then remains the same as in Eq.~\ref{aa:4}, but with $ \tilde{D}$ instead of $D$.

\section{Effective interactions}\label{sec:EffectiveInteractions}

The goal of this section is to derive the effective four-fermion interaction between majority spin-up fermions on the Fermi surface. A single-magnon exchange is insufficient to mediate pairing because it scatters a spin-up electron into a spin-down state, which is gapped in the fully polarized phase. Consequently, we must proceed to the next order in the diagrammatic expansion. First, we construct an effective electron-magnon vertex involving two successive spin-flip scatterings that return the virtual spin-down fermion back to the spin-up Fermi surface (Sec.~\ref{sec:EffectiveEMVertex}). Using this vertex, we then evaluate the effective low-energy interaction between two spin-up electrons mediated by a two-magnon exchange (Sec.~\ref{sec:EffectiveEEInteraction}).
 
\subsection{Effective electron-magnon interaction vertex }\label{sec:EffectiveEMVertex}

\begin{figure}[b]
    \centering
    \includegraphics[width=\linewidth]{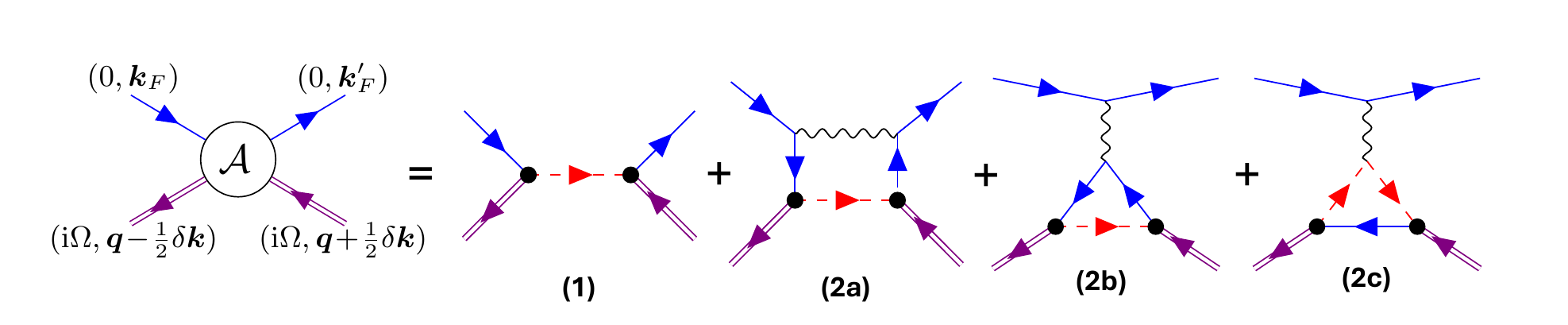}
    \caption{Effective electron-magnon interaction. Solid blue lines denote spin-up electrons; red line dashed lines denote spin-down electrons; purple double lines denote magnon propagator; wavy lines denote bare interaction, and solid dot denotes bare electron-electron-magnon vertex. Diagrams (2a) and (2b) are present in the diagrammatic technique but cancel exactly for a contact interaction as an implementation of the Pauli principle.}
    \label{fig:EffectiveVertex}
\end{figure}
The effective vertex $\Amc(\kbs_F,\kbs_F';\ii\Omega,\qbs)$ for the interaction involving two spin-up electrons and two magnon propagators is shown in Fig.~\ref{fig:EffectiveVertex}. It has two components. One is the direct 2nd order term $4 \Delta^2 U^2  G_{\da}(-\ii\Omega,\sbs-\qbs)$ (Fig.~\ref{fig:EffectiveVertex} (1)).  Another is ``Aslamazov-Larkin" type interaction 
involving an additional $U$ and the convolution of three Green's functions -- two for one spin projection and one for the other.
There are  three such terms in the diagrammatic representation - diagrams (2a)-(2c) in  Fig.~\ref{fig:EffectiveVertex}. 
The  diagrams 2a and 2b involve two spin-up propagators and one spin-down propagator and the bare interaction between spin-down fermions. 
These two diagrams cancel out. This cancellation is a direct manifestation of the Pauli exclusion principle, which states that an onsite Hubbard interaction vanishes for fermions with parallel spins  [In the diagrammatic technique~\cite{agd},  developed for a generic $U(q)$, this interaction is kept, but by Pauli principle, it should not  contribute to a measurable quantity, the cancellation between 2a and 2b shows that this is indeed the case.]   
The diagram (2c) is non-zero and is of the same order as the diagram (1) in this figure.

Combining diagrams (1) and (2c),  we obtain  the total two electron-two magnon interaction vertex in the form  
\begin{equation}\label{eq:EffectiveAVertex}
\begin{split}
    \Amc (\kbs_F,\kbs_F';\ii\Omega,\qbs)
&=
[G_{\da}(-\ii\Omega,\sbs-\qbs) 
 +T(\delta\kbs; \ii\Omega,\qbs)]\\ 
T(\delta\kbs;\ii\Omega,\qbs)& = 
U\int_{\lbs,\ii\omega} G_{\da}(\ii[\omega+\Omega],\lbs-\qbs-\delta\kbs/2)
G_{\da}(\ii[\omega+\Omega],\lbs-\qbs+\delta\kbs/2)
G_{\ua}(\ii\omega,\lbs).
\end{split}
\end{equation}
We placed two spin-up fermions $\kbs$ and $\kbs'$ onto the Fermi surface ($\kbs = \kbs_{F}$,  $\kbs'=\kbs'_{F}$) and set their frequencies to zero.  We also introduced $\sbs := \frac{1}{2}(\kbs_{F}+\kbs_{F}')$, and $\delta\kbs:= \kbs'_F-\kbs_F$. The magnons have momenta $\qbs \pm \delta \kbs/2$ and Matsubara  frequency $\Omega$. 
The vertex component $T(\delta\kbs;\ii\Omega,\qbs)$  is obtained in  App.~\ref{app:Diagram2c}. 
Here we present the result for $ T(\zero; \ii\Omega,\qbs) $. It is: 
\begin{equation}\label{eq:Diagram2c:T}
        T(\zero; \ii\Omega,\qbs) = 
    \frac{mc}{\qbs^2}\left(\frac{1}{\sqrt{1 - 2\frac{\qbs^2}{mc}\frac{2\Delta}{\left(\frac{\qbs^2}{2m}+2\Delta + \ii\Omega \right)^2} }}-1\right)
\end{equation}
The inclusion of the second-order diagrams in $\mathcal{A}$ is necessary to ensure that the electron–magnon interaction vanishes in the limit of vanishing frequency and momentum, i.e., $\mathcal{A}(\mathbf{k}_F,\mathbf{k}_F; i0,\mathbf{0}) = 0$. This condition is required to satisfy the Adler principle \cite{Adler1965,raines2025superconductivity}, which essentially states that the interaction between a Goldstone boson and fermions should vanish at $q=\Omega=0$, otherwise the contribution from a fermionic loop would destroy the Goldstone form of the boson propagator. 

In Sec.~\ref{sec:SCCouplingConstant}, we show (see Eq.~\ref{eq:lambdap}) that the dimensional coupling $\lambda_p$ for the $p-$wave pairing depends on $z = (c-1) E_F/(4 c \Omega_0)$, which we  introduced in Sec.~\ref{sec:Intro}. Near the onset of ferromagnetism, at $c-1 \ll 1$,  this coupling becomes of order one already at large $z \sim 1/(c-1)^{2/3}$. For these large $z$, we find that $\Amc(\kbs_F,\kbs_F';\ii\Omega,\qbs)$ can be approximated as 
\begin{equation}\label{eq:AmcAppprox:q} \Amc(\kbs_F,\kbs_F';\ii\Omega,\qbs) \approx  \frac{1}{4\Delta^2} 
    \frac{\qbs\cdot (\kbs_F+\kbs_F')}{2m} 
\end{equation}
 with corrections that are small in $1/z$. 
 
\subsection{Effective electron-electron interaction}\label{sec:EffectiveEEInteraction}
\begin{figure}[t]
    \centering
    \includegraphics[width=0.5\linewidth]{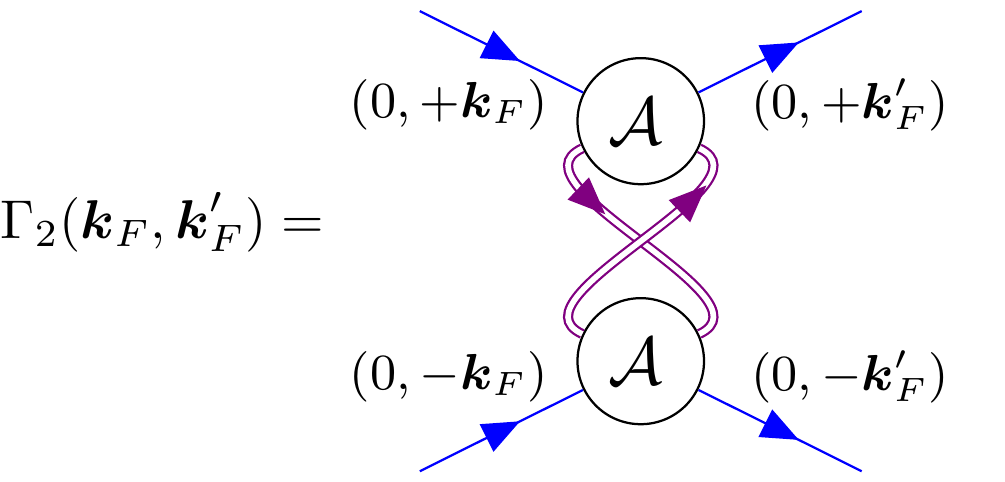}
    \caption{Diagrammatic expansion of the magnon-mediated electron-electron interaction.}
    \label{fig:Gamma2}
\end{figure}

We are now in position to calculate the effective interaction, $\Gamma_2$, between two spin-up electrons on the Fermi surface, mediated by two magnon propagators. This process is represented diagrammatically in Fig.~\ref{fig:Gamma2}. We set the incoming momenta to $ \kbs_F$ and $-\kbs_F$, and the outgoing momenta to be $\kbs_F'$ and $-\kbs_F'$. Combining the results from the previous section, we find
\begin{equation}
   \begin{split}
        \Gamma_2(\kbs_F,\kbs_F')
        &=  -
        [2\Delta U]^2
        \int 
        \frac{\dd\Omega}{2\pi}\frac{\dd^2\qbs}{(2\pi)^2} 
    \Amc(+\kbs_F,+\kbs_F'; \ii\Omega,\qbs) 
    \Amc(-\kbs_F,-\kbs_F'; \ii\Omega,\qbs) 
    {\tilde D}_{\ua\da}(\ii\Omega,\qbs+\tfrac{\delta\kbs}{2})
    {\tilde D}_{\da\ua}(\ii\Omega,\qbs-\tfrac{\delta\kbs}{2})\\
    &=
    \frac{U^2\abs{\kbs_F +\kbs_F'}^2}{32\pi^2m^2[2\Delta]^2} \int_{-\infty}^{+\infty}\dd\Omega \int_0^{q_c}\dd{q}  \left(1 - \frac{q^2}{q^2_c}\right)^2 \int_0^{2\pi}\frac{\dd\theta}{2\pi}
    \frac{2q^3 \sin[2](\theta)}{(B(\ii\Omega) + \alpha[q^2+\delta\kbs^2/4])^2 - \alpha^2 q^2 \abs{\delta \kbs}^2 \cos[2](\theta)},
   \end{split}
\end{equation}
here $\theta$ is the angle of $\qbs$ with respect to $\delta\kbs$. The $\sin[2](\theta)$ in the denominator appears because $(\kbs_F+\kbs_F')$ is orthogonal to $\delta\kbs$. Recall that the momentum cutoff is at $q_c=c\kF$ where the magnon residue vanishes (see Eq.~\ref{aa:5}). Crucially, and in contrast to Ref.~\cite{raines2025superconductivity}, no cutoff is imposed on the frequency integration. 

\begin{figure}
    \centering
    \includegraphics[width=0.5\linewidth]{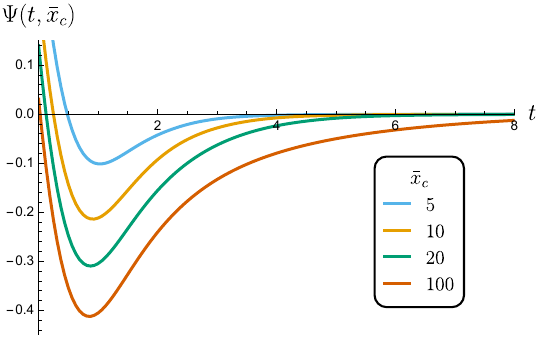}
    \caption{Scaling function    $\Psi(t,\bar{x}_c)$  (Eq.~\ref{eq:Psi of t and bark}) for the effective electron-electron interaction mediated by two magnons. 
    The two variables are  $t=4 z\sin(\phi/2)^2$ and $\bar{x}_c = 4zc \approx 4z$, where $\phi$ is the angle between the incoming and the outgoing momenta on the Fermi surface and $z$ is defined in Eq.~\ref{qqq}. }
    \label{fig:F(t)}
\end{figure}

It is convenient to measure momenta and energy in units of $\sqrt{\Omega_0/\alpha}$ and $\Omega_0$, respectively. We recall that $\Omega_0$ is the energy scale associated with magnetic anisotropy and $\alpha=\frac{c-1}{2mc}$ is the magnon stiffness, see Eq.~\ref{eq:SpinStiffness}. 
We introduce the dimensionless variables 
\begin{equation}
    \bar{x}:= \alpha q^2/\Omega_0; 
    \quad \bar{y}:= \Omega/\Omega_0; 
    \quad \bar{x}_c:= \alpha q_c^2/\Omega_0;
    \quad \bar{B}({\bar y}):=B (\ii {\bar y}\Omega_0)/\Omega_0;
\end{equation}
and the dimensionless parameter
\begin{equation}
z = \frac{\alpha\kF^2}{4\Omega_0} 
= (c-1) \frac{E_F}{4c\Omega_0}.
    \label{qqq}
\end{equation}
Near the onset of ferromagnetism, at $c-1 \ll 1$, $z \approx (c-1) \frac{E_F}{4c\Omega_0}$. In terms of $z$,  $\bar{x}_c =\alpha q^2_c /\Omega_0 =4zc^2$. 
In these new variables, the dimensionless electron-electron interaction can be expressed as 
\begin{gather}
    \Gamma(\phi):=\nu\Gamma_2(\kbs_F,\kbs_F') =
    \frac{c}{c-1}\frac{1}{4\pi z}
       \frac{1+\cos(\phi)}{2}
       \Psi( 4z\sin[2](\phi/2),4zc^2), \label{eq:Gamma2}\\
          \Psi(t, {\bar x}_c)\equiv   
          \int_{0}
          ^{+\infty}\dd \bar{y} \int_0^{\bar{x}_c}   
          {\left(1 - \frac{{\bar x}}{{\bar x}_c}\right)^2 }
          \dd{\bar{x}} \int_0^{2\pi}\frac{\dd\theta}{2\pi}
          \Re\frac{\bar{x} \sin[2](\theta)}{(\bar{B}(\bar{y}) + \bar{x}+{t})^2 - 4\bar{x}{t}\cos[2](\theta)}.\label{eq:Psi of t and bark}
\end{gather}
Here $\phi = \phi_{\kbs_F'}-\phi_{\kbs_F}$ is the angle between $\kbs_F$ and $\kbs_F'$. We evaluated $\Psi(t, {\bar x}_c)$ numerically 
for several values of $\bar{x}_c$. We show the results in  Fig.~\ref{fig:F(t)}. We see that $\bar{x}_c$, $\Psi(t,{\bar x}_c)$ is negative for most $t$ and  has a deep minima at $t\sim 1$. It decays to zero at larger $t$ and changes sign from negative to positive at the smallest $t$.
We also note that the presence of a small but finite $\omega_0$ is essential here. If  $\Omega_0$ is zero,  $B(\Omega) = i \Omega$, i.e., ${\bar B} ({\bar y}) = i {\bar y}$.  The integral over ${\bar y}$ then vanishes because the two poles in the integrand are located in the same half-plane of complex frequency (see App.~\ref{app:no_go} for more discussion on this).  When $\Omega_0$ is finite, the poles at the smallest $\Omega$ are no longer in the same plane and the integral over ${\bar y}$ does not vanish. This consideration also shows that the integral over ${\bar y}$ is confined to ${\bar y} = \Omc(1)$,{ i.e.,} the integral over ${\bar y}$ is infra-red convergent. 

\section{\texorpdfstring{Coupling constant for {$p$-}wave pairing }{Coupling constant for p-wave pairing }}
\label{sec:SCCouplingConstant}

As our interaction is static, we can estimate the critical temperature of the pairing instability by solving the linear equation for the pairing vertex $\Phi(\kbs_F) \sim \expval{\bar{\psi}_{+\kbs_F\ua}\bar{\psi}_{-\kbs_F\ua}}$ (see App.~\ref{app:LinearizedGapEquation} for  details).  
 Collecting the Cooper logarithm from the integration of the product of two fermionic propagators transverse to the Fermi surface, we obtain the integral equation for $\Phi(\kbs_F)$ in the form
\begin{equation}
    \Phi(\kbs_F)  = - L \nu \int \frac{\dd \kbs_F'}{2\pi \kF} \Gamma_2(\kbs_F,\kbs_F')\Phi(\kbs_F').
\end{equation}
Here $L =\log(\EF/T_c)$ is the Cooper logarithm and  $\nu = \frac{m}{2\pi}$ is the density of states.
The effective interaction $\Gamma_2$ is a function of the angle $\phi$ between $\kbs_F$ and $\kbs_F'$, see Eq.~\ref{eq:Gamma2}. 

\begin{figure}
    \centering
    \includegraphics[width=0.5\linewidth]{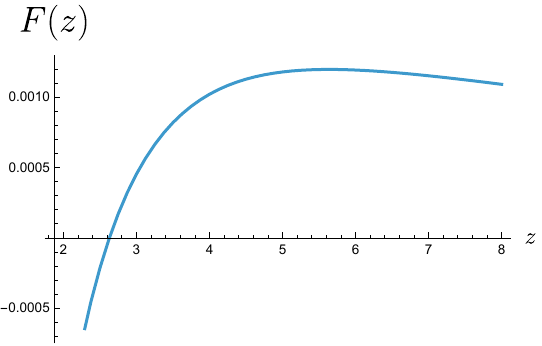}
    \caption{Scaling function $F(z)$ (Eq.~\ref{eq:ScalingFunctionFofz}) that determines the {$p$-}wave coupling constant near the onset of ferromagnetism, at $c-1 \ll 1$. The variable $z$, defined in Eq.~\ref{qqq}, 
    is $z =  (c-1) E_F/(4\Omega_0)$. 
   The function $F(z)$  is  positive (attractive) for $z \geq 1$ and   and decays as $ F(z) \approx 0.1/z^{3/2}$ at large $z$. It becomes negative (repulsive) for smaller $z$.}
    \label{fig:ScalingFunctionFofz}
\end{figure}

Due to Fermi statistics for equal-spin pairing,  $\Phi(\kbs_F)$ must be odd under inversion, {i.e.,} $\Phi(-\kbs_F) = -\Phi(\kbs_F)$. This  selects odd angular momentum channels  $\ell = (2l+1)$, as obvious for polarized fermions. 
Due to rotational invariance,  this integral equation splits between different $\ell$  into a set of independent algebraic equations  for a given $\ell$. The corresponding dimensional pairing coupling constant is 
\begin{equation}
    \lambda_{\ell} = - \int_0^{\pi} \Gamma(\phi)\cos(\ell\phi) \frac{\dd\phi}{\pi}.
\end{equation}
For $p$-wave channel ($\ell=1$), we obtain 
\begin{align}
    \lambda_{p} &= \frac{c}{c-1} F (z) 
\label{eq:lambdap} 
   \end{align}
    where, as before, $z = (c-1) \EF/(4c\Omega_0)$
    and
\begin{equation}\label{eq:ScalingFunctionFofz}
F(z) = -\frac{1}{4\pi z  }\int_0^{\pi} \Psi (4z\sin^2(\phi/2)) \frac{1+\cos(\phi)}{2}\cos(\phi) \frac{\dd{\phi}}{\pi}.
\end{equation}
Eqs.~\ref{eq:lambdap} and~\ref{eq:ScalingFunctionFofz} are the key results of this paper.  
We evaluated $F(z)$ numerically and show the result in Fig.~\ref{fig:ScalingFunctionFofz}.   We see that at $z >1$, where our calculations are under control, $F(z)$ is positive (attractive). It increases with decreasing $z$, approximately as $F(z) = 0.1/z^{3/2}$.  The $p-$wave coupling then becomes unity at $c-1 \approx 0.4 (\Omega_0/E_F)^{3/5}$.
As $4 \Omega_0/E_F < c-1 < 0.4 (\Omega_0/E_F)^{3/5}$,  $\lambda_p$ formally becomes even larger than unity. It is possible that for these $z$, the self-energy correction (mass renormalization) will change $\lambda_p$ into $\lambda_p/(1 + \lambda_p)$, but in any case, for $4 \Omega_0/E_F < c-1 <   0.4 (\Omega_0/E_F)^{3/5}$,  $T_c$ for $p$-wave pairing is not exponentially small and is a fraction of $E_F$, which is the overall scale for magnon-mediated pairing. At larger $c-1$, $\lambda_p$ gets smaller and $T_c$ rapidly (exponentially)  decreases to non-observable values. At smaller $c-1$, when $z \leq 1$, our computations are not under full control because the corrections to $\Amc(\kbs_F,\kbs_F';\ii\Omega,\qbs)$ in Eq.~\ref{eq:AmcAppprox:q} are $\Omc(1)$.
Neglecting these corrections to get an estimate of the pairing strength at $z <1$,  
we find that $F(z<1)$ gets smaller as $z$ decreases and eventually changes sign and becomes repulsive. This implies that $p-$wave superconductivity develops close to the onset of ferromagnetism, but still at a finite distance from a ferromagnetic transition.  

We also remark that at $z\gg 1$, when the relevant $\abs{\delta {\bf k}}/\kF$ are small, the couplings $\lambda_{\ell}$ with larger odd $\ell$, i.e., the couplings in higher angular-momentum channels, are comparable to $\lambda_p$. We verified numerically that this is correct, but the $p$-wave component is always larger. A similar situation holds for the pairing near a nematic instability in a Fermi liquid \cite{Lederer2015,Lederer2017,Klein2018}. There, all $\lambda_{\ell}$ are also comparable, but $\lambda_{0}$ is the largest by magnitude.

\section{Conclusions}
\label{sec:conclusions}

In this manuscript, we analyzed magnon-mediated superconductivity inside the ferromagnetic phase of a 2D itinerant electron system with parabolic dispersion and Hubbard-type interaction within
Hartree-Fock (ladder)  approximation. Within this approximation, a ferromagnetic Stoner instability instantly leads to  a  fully spin polarized half-metal state with Fermi surface only for spin-up fermions. Longitudinal spin fluctuations are absent in such a state, leaving the transverse Goldstone magnons as the only electronic candidate for the pairing glue.  
We derived the effective interaction between spin-up fermions and argued that it necessary includes two magnon propagators. 
We first demonstrated that this interaction vanishes identically for SU(2)-symmetric case because both dynamical magnon propagators in the pairing vertex have poles in the same half-plane of frequency. We then broke SU(2) symmetry down to U(1) by introducing a weak easy-plane anisotropy, which changes  the  magnon spectrum at the smallest frequencies such it has poles in both half-planes  of frequency. We showed that in this case the pairing interaction becomes non-zero.

We set the the anisotropy to be weak, such that the U(1) form of the magnon propagator holds only at frequencies below 
$\Omega_0 \ll E_F$. A naive expectation would be that in this limit the pairing coupling constant is  small  in $\Omega_0/E_F$ and $T_c$ is unobservably small. We showed that this is true deep inside the ferromagnetic phase but not true near its onset at $c =1$.    Specifically, we found an unusual and highly non-trivial behavior of the coupling in the p-wave channel, $\lambda_p$. We obtained this coupling constant as a function of two parameters: the proximity to the onset of ferromagnetism $c-1$ and the dimensionless ratio $\Omega_0/\EF$. At $c-1 = \Omc(1)$, $\lambda_p \propto (\Omega_0/\EF)^{3/5}$ is small, {i.e,} superconductivity is very weak. 
However, at smaller $c-1$, {i.e.,} closer to the magnetic transition, $\lambda_p$ increases and at $(c-1)\sim (\Omega_0/\EF)^{2/5}$ becomes of order unity. We demonstrated that this enhancement occurs within the regime of validity of our computations. In this range,  superconducting $T_c$ is a sizable fraction of the Fermi energy $E_F$. At smaller values of $(c-1)\Omega_0/\EF$ , $\lambda_p$ changes sign and becomes repulsive. This implies that $T_c$ in a ferromagnet becomes non-zero only above a certain distance from the onset of ferromagnetism. 

Ultimately, these results highlight a complex and delicate interplay between spin anisotropy and proximity to a magnetic instability. They demonstrate that robust $p$-wave superconducting inside a ferromagnet can be obtained without relying on multi-band mechanisms or on the existence of both majority and minority spin carriers (a partially  polarized state).

\section*{Acknowledgments}
We would like to thank J. Alicea, E. Berg, Z. Dong, P. Lee, S. Nadj-Perge, A. Thompson, and  A. Young.
The work by AVC was supported by the National Science Foundation grant NSF: DMR-2325357.
The work by VC was supported by the Bill Fine Postdoctoral Fellowship at FTPI. H.L. was supported by the European Research Council (ERC) under the European Unions Horizon 2020 research and innovation program (Grant Agreement No. 101020833). B.A.B. was supported by the Gordon and Betty Moore Foundation through Grant No. GBMF8685 towards the Princeton theory program, the Gordon and Betty Moore Foundation’s EPiQS Initiative (Grant No. GBMF11070), the Global Collaborative Network Grant at Princeton University, the Simons Investigator Grant No. 404513, the NSF-MERSEC (Grant No. MERSEC DMR 2011750), the Simons Collaboration on New Frontiers in Superconductivity (Grant No. SFI-MPS-NFS00006741-01), Princeton Catalysis Initiative (PCI), the Schmidt Foundation at the Princeton University and the National Science Foundation through the AI Research Institutes program Award No. DMR-2433348.

\newpage 
\appendix

\section{Evaluation of diagrams}

In App.~\ref{app:SpinSusceptibilityANDMagnonPole} we evaluate the spin-flip susceptibility (Eq.~\ref{eq:SpinFlipSusceptibilitity} in the main text). In App.~\ref{app:Diagram2c}, we discuss diagram 2c in Fig.~\ref{fig:EffectiveVertex} of the main text. 

\subsection{Spin-spin susceptibility and magnon pole }\label{app:SpinSusceptibilityANDMagnonPole}

In this section, we evaluate the transverse spin susceptibility (or spin-flip bubble) within the ferromagnetic phase. The susceptibility is defined as:
\begin{equation}
\begin{split}\label{eq:Piua:Explicit}
    \Pi_{\ua\da}(\ii\Omega,\qbs) :=& - \int\frac{\dd^2{\lbs} \dd\omega}{(2\pi)^3} G_{\da}(\ii[\omega-\Omega],\lbs-\qbs)G_{\ua}(\ii\omega,\lbs) \\
    = &
    \int\frac{\dd^2{\lbs}}{(2\pi)^2} \frac{\Theta(\kF -\abs{\lbs})}{\ii\Omega + \mu_{\ua}-\mu_{\da} +\frac{(\lbs+\qbs)^2}{2m}- \frac{\lbs^2}{2m}}\\
    = &
    \int_0^{\kF}\frac{l\dd l}{2\pi} \int_0^{2\pi}\frac{\dd\theta}{2\pi} \frac{1}{\ii\Omega + 2\Delta +\frac{\qbs^2}{2m} - l \abs{\qbs} \cos(\theta)/m}
    \end{split}
\end{equation}
The angular integration can be evaluate using the following definite integral
\begin{equation}
    \int_0^{2\pi} \frac{1}{1-a\cos(\theta)} \frac{\dd{\theta}}{2\pi} = \frac{1}{\sqrt{1-a^2}} \quad \text{for} \abs{a}<1].
\end{equation}
Performing a subsequent elementary integration over the kinetic energy $\varepsilon = \frac{l^2}{2m}$ yields
\begin{equation}\label{eq:Piua:Final}
    \Pi_{\ua\da}(\ii\Omega,\qbs) :=\frac{m}{2\pi} \frac{m}{q^2}\left(\frac{q^2}{2m}+2\Delta + \ii\Omega  - \sqrt{\left(\frac{q^2}{2m}+2\Delta + \ii\Omega \right)^2 - \frac{4\Delta}{c}\frac{q^2}{m}}\right).
\end{equation}
where we have set $q := \abs{\qbs}$.

The magnon pole is determined by identifying the zeros of the denominator of the spin-flip susceptibility (Eq.~\ref{eq:chi:ud}):
\begin{equation}
    1 - U \Pi_{\ua\da}(\ii\Omega,\qbs) = 0.
\end{equation}
to locate the pole, we use the ansatz 
$\ii\Omega = - E_{\rm{m}}(\qbs)$, with 
\begin{equation}
    E_{\rm{m}}(\qbs) = \frac{\qbs^2}{2m}\left(1- \frac{1}{c}\right),
\end{equation}
Substituting this ansatz into the pole condition yields:
\begin{equation}
 1-U\Pi_{\ua\da}(-E_{\rm{m}}(\qbs),\qbs) = 1- 
 \frac{2mc}{q^2} \min( \frac{q^2}{2mc},2\Delta).
\end{equation}
This confirms that the magnon pole is strictly located at $\ii \Omega = - E_{\rm{m}}(\qbs) = - \alpha\qbs^2$, where $\alpha = \frac{c-1}{2m c}$, as long as $\qbs^2 < 4mc\Delta$.

Next, we calculate the residue of the magnon pole
\begin{equation}\label{eq:MagnonWeight}
   Z_{\rm{m}}(\qbs)= \lim_{\ii \Omega \to - \alpha\qbs^2} \frac{\ii\Omega+\alpha\qbs^2}{1- U \Pi_{\ua\da}(\ii\Omega,\qbs)} = 2\Delta\left(1 - \frac{\qbs^2}{4mc \Delta}\right).
\end{equation}
This result demonstrates that the spectral weight vanishes continuously as the magnon pole disappears. 

We thus approximate 
\begin{equation}
    \chi_{\ua\da}(\ii\Omega,\qbs) \approx \frac{Z(\qbs)}{\ii\Omega+\alpha\qbs^2}.
\end{equation}
This expression agrees with Ref.~\cite{raines2025superconductivity} after setting $Z$ to unity.

\subsection{Diagram 2c}\label{app:Diagram2c}

To evaluate the vertex correction from diagram 2c in Fig.~\ref{fig:EffectiveVertex}, we define the shifted external momenta as $\qbs_\pm = \qbs \pm \delta \kbs/2$. Following the definition of the effective vertex in Eq.~\ref{eq:EffectiveAVertex}, the contribution of the diagram can be explicitly written as
\begin{equation}
      \begin{split}
          T(\delta\kbs;\ii\Omega,\qbs) 
          &= 
    -U\int\frac{\dd^2{\lbs} \dd\omega}{(2\pi)^3} G_{\da}(\ii[\omega+\Omega],\lbs-\qbs_-)
    G_{\da}(\ii[\omega-\Omega],\lbs-\qbs_+)
    G_{\ua}(\ii\omega,\lbs)\\
      &= 
        U\int\frac{\dd^2{\lbs}}{(2\pi)^2} \Theta(\kF- \abs{\lbs}) 
        \frac{1}{\ii\Omega + 2\Delta + (\frac{(\lbs-\qbs_+)^2}{2m} - \frac{\lbs^2}{2m})}
        \frac{1}{\ii\Omega + 2\Delta + (\frac{(\lbs-\qbs_-)^2}{2m} - \frac{\lbs^2}{2m})},
      \end{split}
\end{equation}
where $\Theta$ is the Heaviside theta function.

We first examine the limit of vanishing relative momentum, setting $\delta\kbs =0$. In this collinear kinematic regime, the vertex correction can be expressed in terms of the spin-flip bubble: 
\begin{equation}
\begin{split}
    T(\zero;\ii\Omega,\qbs) 
    &= U \int \frac{\dd^2\lbs}{(2\pi)^2}\Theta(\kF-\abs{\lbs})  \ii\pdv{ \Omega} \frac{1}{\ii\Omega  + 2\Delta +(\epsilon_{\lbs+\qbs}-\epsilon_{\lbs})}\\
    &= U \ii \pdv{\Omega}\Pi_{\ua\da}(\ii\Omega,\qbs)\\
    &=- \frac{Um }{2\pi}
    \frac{m}{\qbs^2}\left(1 - \frac{1}{\sqrt{1 - 2\frac{{\qbs}^2}{mc}\frac{2\Delta}{\left(\frac{\qbs^2}{2m}+2\Delta + \ii\Omega \right)^2} }}\right),
\end{split}
\end{equation}
where we used Eqs.~\ref{eq:Piua:Explicit} and~\ref{eq:Piua:Final}. Recalling that $c= Um/2\pi$, we find
\begin{equation}
    T(\zero;0,\zero)= \frac{1}{2\Delta},
\end{equation}
in agreement with Ref.~\cite{raines2025superconductivity}.

Alternatively, we can calculate evaluate the static ($\Omega=0$), long-wavelength ($\qbs=\zero$) limit while maintaining a finite momentum $\delta\kbs$. 
\begin{equation}
   \begin{split}
        T(\delta\kbs;0,\zero) &= U\int\frac{
        \dd^2\lbs}{(2\pi)^2} \frac{1}{2\Delta +\frac{[\delta\kbs]^2}{8m} - \frac{1}{2m}\delta\kbs \cdot \lbs }
        \frac{1}{2\Delta +\frac{[\delta\kbs]^2}{8m} + \frac{1}{2m}\delta\kbs \cdot \lbs }\Theta(\kF-\abs{\lbs})\\
        &= \frac{U}{(2\pi)^2}\int_{0}^{\kF} \int_0^{2\pi} \frac{l}{\left[2\Delta + \frac{[\delta\kbs]^2}{8m}\right]^2- [\frac{\abs{\delta\kbs}l}{2m}\cos(\theta)]^2}
        \dd{l}\dd{\theta}\\
        &=\frac{U}{2\pi}\int_0^{\kF} \frac{l}{(2\Delta +\frac{\abs{\delta\kbs}^2}{8m})^2\sqrt{1 - \frac{[\delta\kbs]^2}{2m}\frac{ l^2}{2m} \frac{1}{(2\Delta+\frac{\delta\kbs^2}{8m})^2}}}\dd{l}\\
        &=\frac{mU}{2\pi}\frac{2}{\frac{\delta\kbs^2}{2m}} \left[1-\sqrt{1- \frac{\delta\kbs^2}{2m}\frac{\kF^2}{2m} \frac{1}{\left(2\Delta+ \frac{\delta\kbs^2}{8m}\right)^2}}\right]\\
        &=\frac{4mc}{\delta\kbs^2} \left[1-\sqrt{1- 2\frac{\delta\kbs^2}{4mc} \frac{2\Delta}{\left(2\Delta+ \frac{\delta\kbs^2}{8m}\right)^2}}\right]\\
        &\approx \frac{1}{2\Delta} - \frac{(2c-1) \delta \kbs^2}{32 c m\Delta^2} +\Omc(\delta\kbs^4).
   \end{split}
\end{equation}
To perform the angular integration in the second line, we have utilized the definite integral identity:
\begin{equation}
    \int_0^{2\pi} \frac{1}{A-B\cos[2](\theta)}\dd{\theta} = \frac{2\pi}{A\sqrt{(1-B/A)}};\quad \text{for} \abs{B/A}<1.
\end{equation}
The subsequent radial integration over l is straightforwardly evaluated by executing the algebraic substitution:
\begin{equation}
L:= \frac{\delta\kbs^2}{2m}\frac{l^2}{2m}\frac{1}{(2\Delta+\frac{\delta\kbs^2}{8m})^2}\end{equation}
and $\int (1-L)^{-1/2}\dd{L} = -2 (1-L)^{1/2}$. In the secont to last step, we used the relation $\frac{\kF^2}{2m} = 2\mu_0 =\frac{2\Delta}{c}$.

To capture the low-energy behavior systematically, we perform a multi-variable Taylor expansion of $T(\delta \kbs,\ii\Omega,\qbs)$ to leading order in $\frac{\qbs^2}{2m \Delta},\frac{\delta\kbs^2}{2m\Delta},\Omega/\Delta$:
\begin{equation}
    \begin{split}
            T(\delta \kbs,\ii\Omega,\qbs) 
            &=U\int\frac{\dd^2{\lbs}}{(2\pi)^2} \Theta(\kF- \abs{\lbs}) 
        \frac{1}{2\Delta -\frac{\lbs\cdot\qbs_+}{m}+[\ii\Omega + \frac{\qbs_+^2}{2m}]}
        \frac{1}{2\Delta -\frac{\lbs\cdot\qbs_-}{m}+[\ii\Omega + \frac{\qbs_-^2}{2m}]}
        \\
        &\approx \frac{U}{[2\Delta]^2}\int\frac{\dd^2{\lbs}}{(2\pi)^2} \Theta(\kF- \abs{\lbs}) 
        \left(1+ \frac{\lbs\cdot\qbs_+}{2m\Delta}+\left[\frac{\lbs\cdot\qbs_+}{2m\Delta}\right]^2 - \frac{1}{2\Delta}\left[\ii\Omega+\frac{\qbs^2_+}{2m}\right]\right)\\
        &\hspace{3cm} \times 
        \left(1+ \frac{\lbs\cdot\qbs_-}{2m\Delta} +\left[\frac{\lbs\cdot\qbs_-}{2m\Delta}\right]^2
        - \frac{1}{2\Delta}\left[\ii\Omega+\frac{\qbs^2_-}{2m}\right]\right)
        \\
&\approx \frac{1}{2\Delta}\left(1 -\frac{1}{2\Delta}\left[ 2\ii\Omega +  \frac{\qbs_+^2}{2m}(1-\tfrac{1}{c})+ \frac{\qbs_-^2}{2m}(1-\tfrac{1}{c}) - \frac{\qbs_+\cdot\qbs_-}{2m c} \right]\right)\\
& = 
\frac{1}{2\Delta}\left(1 -\frac{1}{2\Delta}\left[ 2\ii\Omega +  \frac{\qbs^2}{2m}(2-\tfrac{3
}{c})+ \frac{\delta\kbs^2}{8m}(2-\tfrac{1}{c})\right]\right)
    \end{split}
\end{equation}
where we used the following integrals
\begin{equation}
   \begin{split}
           \int \frac{\dd^2\lbs}{(2\pi)^2} \Theta(\kF-\abs{\lbs})  &= \frac{\kF^2}{4\pi} = \frac{2\Delta}{U}\\
        \int \frac{\dd^2\lbs}{(2\pi)^2} \Theta(\kF-\abs{\lbs}) [\lbs\cdot \abss]&= 0\\
        \int \frac{\dd^2\lbs}{(2\pi)^2} \Theta(\kF-\abs{\lbs}) \frac{[\lbs\cdot \abss][\lbs\cdot \bbs]}{2m} &= \frac{\kF^4}{32m\pi}\abss\cdot\bbs = \frac{2\Delta}{U}\frac{\Delta}{2c}\abss\cdot\bbs  
   \end{split}
\end{equation}
where $\abss$ and $\bbs$ are two dimensional vectors. 

Crucially, across all the regimes examined, the vertex correction consistently approaches $T\to 1/(2\Delta)$, scaling inversely with the ferromagnetic order parameter. This behavior is expected to satisfy the Adler principle, ensuring the proper decoupling of the Goldstone modes (magnons) in the long-wavelength limit.

\section{Linearized gap equation}\label{app:LinearizedGapEquation}

For the sake of completeness, we re-derive how to related $T_c$ with the `pairing coupling constant' for a static interaction $\Gamma_2(\kbs,\kbs')$ (see Chapter 54 of Ref.~\cite{LandauLifshitzVol9} for  the derivation in three dimensions). We start with the linearized gap equation 
\begin{equation}
    \Phi(\kbs) = -T\sum_{\ii\omega_{n}} \int_{\kbs'}\Gamma_2(\kbs,\kbs')\Phi(\kbs') 
    G(+\ii\omega,\kbs')
    G(-\ii\omega,-\kbs')
    = -\int_{\kbs'}\Gamma_2(\kbs,\kbs') \Phi(\kbs') \frac{\tanh(\frac{\epsilon_{\kbs'}}{2T})}{2\epsilon_{\kbs'}}
\end{equation}
where $\Phi(\kbs)$ is the gap. We now assume that the gap equation is dominated by $\kbs$ near the Fermi surface. We then perform the momentum integration perpendicular to the Fermi surface. This achieved by the change of variables $\kbs \to (\epsilon,\theta_{\kbs})$, under which the integration measure transforms as $\frac{\dd^2{\kbs}}{(2\pi)^2}\to \nu\frac{\dd{\theta_{\kbs}}}{2\pi}\dd{\epsilon_{\kbs}}$, where $\nu$ is the density of states at the Fermi energy. We then perform the integration over $\epsilon_{\kbs}$ by neglecting the dependence of $\Gamma_2$ and $\Phi$ on this variable. We start by integrating by parts
\begin{equation}
   \begin{split}
       \int_0^X \frac{\tanh(x)}{x}\dd{x} &= \tanh(X)\log(X) - \int_0^X \frac{\log(x)}{\cosh[2](x)}\dd{x}\\
       &=
       \log(X) - \int_0^{\infty}\frac{\log(x)}{\cosh[2](x)}\dd{x} + \left(\log(X) \frac{-2}{\ee^{2X}+1} + \int_X^{\infty} \frac{\log(x)}{\cosh[2](x)}\dd{x}\right)
   \end{split} 
\end{equation}
The terms in parenthesis decay at least as $\log(X)\ee^{-2X}$. Therefore, up this accuracy
\begin{equation}
    \int_0^{X}\frac{\tanh(x)}{x}\dd{x} \approx \log(\frac{4\ee^{\gamma_{\rm{E}}}X}{\pi}),
\end{equation}
where we used a numerical approximation for the exact integral
\begin{equation}
    \int_0^\infty \frac{\log(x)}{\cosh(x)}\dd{x} = \log(\frac{\pi}{4}\ee^{-\gamma_{\rm{E}}}) \approx \log(0.441);
\end{equation}
where $\gamma_{\rm{E}}\approx 0.577216$ is Euler's constant.

Thus, to logarithmic accuracy, the gap equation becomes
\begin{equation}
    \Phi(\kbs_F) = -\nu \log(\frac{2\ee^{\gamma_{\rm{E}}}W}{\pi T})\int_0^{2\pi} \frac{ \dd{\theta_{\kbs_F'}}}{2\pi}\Gamma_2(\kbs_F,\kbs_F')\Phi(\kbs_F'),
\end{equation}
where $W$ is a energy cut-off, and the subscript $F$ in $\kbs_F$ and $\kbs_F'$ means that these momenta are on the Fermi surface.

The gap equation is invariant continuous rotations ($(\theta_{\kbs_F},\theta_{\kbs_F}') \to (\theta_{\kbs_F}+\delta\theta,\theta_{\kbs_F}'+\delta\theta)$ for arbitrary $\delta\theta$). Thus, we can decompose the gap equation into angular momentum channels 
\begin{equation}
    \Phi(\kbs) = \Phi_{\ell_z}\ee^{\ii\theta_{\kbs} \ell_z}.
\end{equation}

Then, the linearized gap equation reduces to 
\begin{equation}
    \Phi_{\ell_z}\left[1+ \log(\frac{2\ee^{\gamma_{\rm{E}}}W}{\pi T}) \left(\int_0^{2\pi} \frac{\dd{\phi}}{2\pi}\Gamma(\phi)\ee^{\ii\phi\ell_z}\right)\right]=0.
\end{equation}
where $\Gamma=\nu \Gamma_2$ is the dimensionless electron-electron interaction, and $\phi$ is the angle between $\kbs_{F}'$ and $\kbs_{F}$. We thus define the coupling constant as 
\begin{equation}
    \lambda_{\ell_z}\equiv -\int_0^{2\pi} \frac{\dd{\phi}}{2\pi}\Gamma(\phi)\ee^{\ii\phi\ell_z},
\end{equation}
so that the critical temperature for the channel with angular momentum $\ell_z$ is
\begin{equation}
    T_{c,\ell_z}  =\frac{2\ee^{\gamma_{\rm{E}}}}{\pi} W\exp(-1/\lambda_{\ell_z}).
\end{equation}

If $\Gamma(\phi)=\Gamma(-\phi)$, as happens for the model considered in the main text, there is a degeneracy between positive and negative angular momenta channels 
\begin{equation}
    \lambda_{\ell_z}
    =
    -\int_0^{2\pi} \frac{\dd{\phi}}{2\pi}\Gamma(\phi)\ee^{\ii\phi\ell_z}
    =
    -\int_0^{2\pi} \frac{\dd{\phi}}{2\pi}\Gamma(-\phi)\ee^{\ii(-\phi)\ell_z} = 
    \lambda_{-\ell_z}.
\end{equation}
We can thus simply write
\begin{equation}
    \lambda_{\ell_z} = -\int_0^{2\pi} \frac{\dd{\phi}}{2\pi}\Gamma(\phi)\cos(\phi\ell_z).
\end{equation}

\section{\texorpdfstring{No go argument from $\da$ loops}{No go argument from down-arrow loops}} 
\label{app:no_go} 
  
An alternative approach to demonstrate that the pairing interaction, mediated by two-magnon exchange, vanishes for a spin-isotropic system without an upper cutoff on frequency integration involves a direct inspection of the individual diagrams contributing to $\Gamma_2$. 

Each diagram contributing to $\Gamma_2$ contains a closed loop consisting entirely of spin-down electron propagators. By routing the internal variables such that a single loop frequency flows along this minority-spin loop, the corresponding frequency integration vanishes identically. This vanishing is a direct consequence of the full spin-polarization, which restrict all of the spin-down Green's functions to lie on the same half of the frequency complex plane. 

This perspective also clarifies the essence of our phenomenological approach: magnetic anisotropy introduces interaction vertices between spin-up and spin-down electrons. Once these sectors are coupled, the argument for vanishing fermionic loops simply no longer applies.

Strictly speaking, this diagrammatic pole argument is only valid at zero temperature ($T=0$). We leave a more thorough investigation of $\Gamma_2$ at finite temperatures, as well as the exact nature of its asymptotic behavior as $T\to 0$, for future work. In this low-temperature regime, analyzing the linearized gap equation may prove insufficient; as demonstrated in previous work on singular pairing interactions (e.g., Ref.~\cite{Chubukov2003}), non-trivial solutions may instead emerge dynamically from the full non-linear gap equation.

\bibliographystyle{apsrev4-2}
\bibliography{References}

\end{document}